\documentclass[12pt,letterpaper]{article}
\usepackage[margin=1in]{geometry}
\usepackage{color}
\usepackage{comment}
\usepackage[utf8]{inputenc}
\usepackage[numbers]{natbib}
\usepackage{graphicx}
\usepackage{amsmath}
\usepackage{amssymb}
\usepackage{lineno}
\usepackage{hyperref}
\setcitestyle{square}

\usepackage{sectsty}
\sectionfont{\large}
\subsectionfont{\normalsize}
\subsubsectionfont{\normalsize}
\paragraphfont{\normalsize}
\usepackage[compact]{titlesec}

\begin{document}

\pagenumbering{gobble}
\huge
\begin{center}
Astro2020 Science White Paper
\vspace{0.1in}

Wide-field Multi-object Spectroscopy to Enhance Dark Energy Science from LSST
\end{center}
\normalsize

\noindent\textbf{Thematic Areas:} Cosmology and Fundamental Physics\\

\noindent\textbf{Principal (corresponding) author:}\\
Name: Rachel Mandelbaum\\
Institution: Carnegie Mellon University\\
Email: rmandelb@andrew.cmu.edu \\
Phone: 412-268-1714\\

\noindent\textbf{Co-authors:} J.~Blazek (SNSF Ambizione, EPFL; CCAPP, Ohio State U.), N.~E.~Chisari (Oxford), T.~Collett (IoCG, Portsmouth), L.~Galbany (U. Pittsburgh, PITT PACC), E.~Gawiser (Rutgers), R.~A.~Hlo\v{z}ek (Toronto), A.~G.~Kim (LBNL), C.~D.~Leonard (CMU), M.~Lochner (African Institute for Mathematical Sciences, South African Radio Astronomy Observatory), J.~A.~Newman (U. Pittsburgh, PITT PACC), D.~J.~Perrefort (U. Pittsburgh, PITT PACC), S.~J.~Schmidt (UC Davis), S.~Singh (UC Berkeley, BCCP), and M.~Sullivan (Southampton), for the LSST Dark Energy Science Collaboration\\

\noindent\textbf{Abstract:} LSST will open new vistas for cosmology in the next decade, but it cannot reach its full potential without data from other telescopes.  Cosmological constraints can be greatly enhanced using wide-field ($>20$ deg$^2$ total survey area), highly-multiplexed optical and near-infrared multi-object spectroscopy (MOS) on 4--15m telescopes. This could come in the form of suitably-designed large surveys and/or community access to add new targets to existing projects. 
First, photometric redshifts can be calibrated with high precision using cross-correlations of photometric samples against spectroscopic samples at $0 < z < 3$ that span thousands of sq. deg.  Cross-correlations of faint LSST objects and lensing maps with these spectroscopic samples can also improve weak lensing cosmology by constraining intrinsic alignment systematics, and will also provide new tests of modified gravity theories.  Large samples of LSST strong lens systems and supernovae can be studied most efficiently by piggybacking on spectroscopic surveys covering as much of the LSST extragalactic footprint as possible (up to $\sim20,000$ square degrees). Finally, redshifts can be measured efficiently for a high fraction of the supernovae in the LSST Deep Drilling Fields (DDFs) by targeting their hosts with wide-field spectrographs.  Targeting distant galaxies, supernovae, and strong lens systems over wide areas in extended surveys with (e.g.) DESI or MSE in the northern portion of the LSST footprint or 4MOST in the south could realize many of these gains; DESI, 4MOST, Subaru/PFS, or MSE would all be well-suited for DDF surveys.  The most efficient solution would be a new wide-field, highly-multiplexed spectroscopic instrument in the southern hemisphere with $>6$m aperture. In two companion white papers we present gains from deep, small-area MOS and from single-target imaging and spectroscopy.  

\newpage
\pagenumbering{arabic}

\section{Introduction}



The Large Synoptic Survey Telescope (LSST; \cite{Ivezic08,2009arXiv0912.0201L}) will transform our knowledge of cosmology over the years 2023--2033, constraining fundamental cosmological physics using a variety of probes.  However, the baseline dark energy analyses that will be carried out by the LSST Dark Energy Science Collaboration (DESC) will require additional data from other ground-based facilities to reduce systematic uncertainties and realize the full potential of LSST \cite{descsrd}.  
In this white paper, we describe how community access to {\bf wide-field ($>20$ deg$^2$ total survey area), highly-multiplexed} optical and near-infrared multi-object spectroscopy (MOS) on 4--15m telescopes can increase the return from the LSST dataset and unlock additional scientific opportunities for every major cosmological probe. 
%
In two companion white papers, we describe the gains for LSST cosmology from community access to deep multi-object spectroscopy on larger telescopes and from follow-up single-target imaging and spectroscopy of supernovae and strong lens systems \cite{deep,single}.  

\section{Photo-{\em z} Calibration via Spectroscopic Cross-Correlations}
\label{sec:crosscorr}


Photometric redshifts (photo-$z$'s) are a critical tool for DESC, as cosmological tests rely on determining the behavior of some quantity with redshift, but we cannot measure spectroscopic redshifts (spec-$z$'s) for the large numbers of objects detected in LSST imaging.  
If photo-$z$ estimates have some undetected systematic bias, dark energy inference can be catastrophically affected (see, e.g., \cite{hearin_etal10,descsrd}); as a result, photo-$z$'s are both a critical tool and a major source of systematic uncertainty in cosmological analyses. 
Without a comprehensive knowledge of galaxy evolution, the only way to reduce photo-$z$ errors and characterize biases is with galaxies with robust spectroscopic redshift measurements.  We follow \cite{specneeds} in dividing the uses of spec-$z$'s into two  classes, ``training'' and ``calibration.'' 
 {\bf Training} is the use of samples with known $z$ to improve photo-$z$ algorithms and  reduce  {\em random} errors on individual objects' photo-$z$'s,  
%
while {\bf calibration} is the problem of determining  the true overall $z$ distribution of a galaxy sample.  Miscalibration will lead to {\em systematic} errors in photo-$z$'s and resulting analyses \cite{2006JCAP...08..008Z,2006astro.ph..5536K,tysonconf,hearin_etal10}.  Needs for MOS surveys for training are discussed in a companion paper \cite{deep}.

Photo-$z$ calibration requirements for LSST are extremely stringent; e.g., the mean and standard deviation of redshift distributions must be known to $\sim 0.002(1+z)$ \cite{descsrd}.  
Based on the systematic incompleteness of existing deep surveys, we proceed under the assumption \citep[e.g., as applied in][]{specneeds} that direct calibration via a large, representative spec-$z$ sample will not be possible given the depth of LSST.
However, methods based upon cross-correlating the locations of LSST galaxies with spec-$z$ samples 
can provide an accurate photo-$z$ calibration for dark energy applications  \cite{2008ApJ...684...88N}. These techniques exploit the fact that all galaxies cluster together on large scales, so those objects with spec-$z$'s cluster on the sky only with the subset of galaxies in other samples that are at nearly the same redshift. 
We can use this to determine the true redshift distribution of objects in an unknown sample \citep[e.g.,][]{Schmidt:2013sba,Rahman:2014lfa} as input to dark energy measurements.  Precise calibration only requires that the spectroscopic samples span the redshift range of the LSST objects and cover a wide area of sky.  

The DESI survey \cite{2016arXiv161100036D} which will begin in 2020 should meet the sample size and redshift range requirements \cite{specneeds} for this application: it will overlap with at least 4000~deg$^2$ of the LSST footprint, obtaining more than 5 million redshifts of galaxies from $0<z<1.6$ and quasars at $z\lesssim$ 3--4.  
However, because DESI is located at the northern end of the LSST 
footprint, there is a risk of the LSST photometry -- and hence the photo-$z$'s to be calibrated -- being systematically different in this region (e.g., due to difficulties in correcting for the atypical distribution of airmasses). 
As a result, greater southern coverage is extremely desirable.  The 4MOST survey, which will use a dedicated wide-field multi-object spectrograph on the VISTA telescope at Paranal Observatory \cite{4most}, plans to obtain samples with DESI-like target selection (but lower number density) over $\sim$1000 deg$^2$ for Emission Line Galaxies (ELGs) and $\sim$7500 deg$^2$ for Luminous Red Galaxies (LRGs) and quasars; these areas may overlap with DESI.  It would thus be valuable to enlarge the samples available for cross-correlation calibration in the southern part of the LSST footprint; this would require new survey data in regions of sky not covered by DESI.




\section{Characterizing Intrinsic Alignments via Cross-Correlations}

Intrinsic correlations between shapes of galaxies that are physically near each other (``intrinsic alignments’’  or IA) are a known contaminant to weak gravitational lensing (WL) \citep{Brown02}. If not accounted for, their presence can generate significant biases in cosmological WL analyses  \citep{Kirk12, Krause16, Blazek17}. 
To reduce final statistical uncertainties, IA modeling should have as few parameters as possible with strong priors on each. 
%
%
Wide-field MOS will enable measuring the cross-correlation between positions of bright galaxies with spectroscopy and intrinsic shapes of fainter galaxies used for lensing analyses, constraining IA models with greater precision than possible with photometric data alone. Such measurements can place upper limits on IA or constrain parameters associated with IA models \citep[e.g.,][]{2014MNRAS.445..726C,Singh14,Johnston18}. 
Furthermore, cross-correlations using the spec-$z$ sample will allow us to break degeneracies between IA parameters and photo-$z$ uncertainties. 
%
%
The LSST-DESI overlap should enable measurements which greatly surpass what is currently possible, especially if northern coverage by LSST is increased (as proposed in \citep{BigSky}).  4MOST galaxy samples will further enhance these analyses.  However, larger samples over wider areas would yield better constraints on IA models, improving S/N in IA measurements by up to $\sim2\times$ if samples are expanded to cover the full LSST extragalactic footprint. Smaller-field surveys of fainter galaxies will provide complementary constraints on IA, as discussed in a companion paper \cite{deep}. 

\section{Testing General Relativity on Cosmological Scales}

Galaxies and WL maps trace the same underlying matter field. Galaxies are biased tracers of matter but 
with spec-$z$'s can  provide full three-dimensional information, while 
lensing maps are unbiased tracers that 
mainly provide information about projected density. Cross-correlating galaxies with lensing maps can provide very fine tomographic information about matter clustering independent of galaxy bias \citep[see, e.g.,][]{Baldauf2010,Mandelbaum2013}. 
Combining cross-correlations between galaxy density and lensing with measurements of redshit-space distortions in galaxy clustering allows for tests of gravity on cosmological scales. Previous  work has defined a  statistic $E_G$ which combines information from these quantities and is sensitive to deviations from General Relativity \cite{Reyes2010, Zhang2007}. 
A sample of galaxies with spec-$z$'s that overlaps the LSST WL sample is required to measure $E_G$. 
DESI and 4MOST should provide such overlapping samples via both LRGs with $\bar{z} \approx 0.77$ and ELGs with $\bar{z}\approx 1.0$.  The combined samples of DESI and 4MOST ELGs and LRGs should each enable a determination of $E_G$ to $\sim 0.004$, roughly 10 times more precise than current constraints.  
Enlarging the northern LSST footprint (as proposed in Ref.~ \cite{BigSky}) would reduce errors by $\sim25\%$ more; greater improvements are possible with enlarged spectroscopic samples in the southern LSST footprint.

\section{Characterizing Strong Lensing Systems from LSST}

LSST will discover $\sim 100,000$ strong gravitational lenses \citep{Collett15}, 100 times more than are currently known. This drastic change opens new opportunities to precisely constrain dark energy with lensed quasars \citep{Bonvin17}, lensed supernovae \citep{Goldstein18}, lenses with multiple background sources \citep{Collett14}, and lenses with a spectroscopic velocity dispersion \citep{Grillo08}, as well as by using strong lensing as a calibration tool for weak lensing shear \citep{Birrer18}. Each of these science cases requires redshifts for lens and source, both to confirm candidates as lenses and to convert lensing-derived quantities into cosmologically meaningful constraints. The LSST lenses will be uniformly distributed across the LSST footprint, with typical $r$-band magnitudes of $20.1 \pm 2.4$ for the lens and $23.7 \pm 0.7$ for the source. As a result, many lenses are bright enough for redshift measurements via targeted fibers within very wide-area surveys, enabling identification of the systems best suited for follow-up.  Our needs for detailed follow-up once the best systems are found are described in a companion paper \cite{single}. 

\section{Spectroscopy for Supernova Cosmology}

Type Ia supernovae (SNe Ia) provide a mature and well-exploited probe of the accelerating universe (e.g., \cite{2018arXiv181102374D}), and their use as standardizable candles is an immediate route to measuring the equation of state of dark energy. LSST, for example, could assemble around 100,000 SNe Ia to $z=1$, giving unprecedented insight into the expansion history of the universe. A major systematic uncertainty will be the photometric classification and redshift measurement of the supernova detections. Wide-field spectroscopy can exploit the fact that wherever a follow-up facility points in the extragalactic sky, there will be known time-variable sources, including both recently discovered transients and older, now-faded events.

Spectroscopy serves two main goals over both the typical LSST "Wide, Fast, Deep" (WFD) footprint and the more frequently-observed Deep Drilling Fields (DDFs).  
The first is the classification of live SNe and the construction of optimized training samples for photometric classifiers to assemble the next generation of SN Ia cosmological samples. 
Even the most advanced machine-learning classification techniques are fundamentally dependent on large, homogenous and representative training sets \citep{2016ApJS..225...31L}. 
%
The second goal is to obtain spec-$z$'s for host galaxies of SNe that have faded away. While conventional SNIa cosmology analyses rely on spectroscopic follow-up of all the SNe, new analyses \citep[e.g.][]{jones2018,campbell2013, hlozek2012} show that it is possible to take advantage of even larger samples of SNe after obtaining spec-$z$'s of their host galaxies. Targeted campaigns of multi-object spectroscopy of supernovae and their hosts in LSST Deep Drilling Fields (DDFs), combined with the assignment of fibers to such objects in larger surveys covering the wide LSST footprint, can provide the redshifts needed for most LSST SN Ia cosmology studies. 
More detailed investigation of a smaller set of supernovae will remain valuable, however; needs for such data are described in a companion paper \cite{single}.

In addition to tests of the background expansion, supernovae provide constraints on the growth of structure through their peculiar velocities. 
Spatial correlations in the peculiar velocities of the supernova sample result from the growth of structure, and the wide field of LSST 
will provide 
a calibrated sample over sufficiently large sky areas to make supernovae a competitive probe of growth, particularly at low redshift (c.f.~\cite{Kim19}).

\subsection{Supernovae and their Hosts in LSST Deep Drilling Fields (DDFs)}



Due to the higher-quality light curves provided by more frequent and deeper photometry, the best-characterized and deepest LSST SN samples will come from the DDFs.  Because of the comparatively small area covered by these fields ($\sim50$ deg$^2$ total) and their low surface density, it is feasible to obtain long exposures on all DDF live supernovae or their hosts using  wide-field spectrographs (e.g., DESI or Subaru/PFS in the North and 4MOST in the South); c.f.~\cite{kavli}. This approach has proven very successful in the OzDES survey \cite{ozdes}, and will be employed by the 4MOST/TiDES survey for LSST supernovae \cite{tides}. 
In addition to the cosmological measurements enabled by DDF SN and host redshifts, the resulting dataset should be an extremely valuable source of high S/N templates for light curves to be used in cosmology for supernovae that have LSST data but no spectroscopy, and will also provide a wealth of training data for photometric classification of transients.


\subsection{Supernovae and their Hosts in the Wide, Fast, Deep (WFD) Survey}

\begin{figure}[htbp!]
\begin{center}
\includegraphics[width=1.0\textwidth,trim=0cm 14.5cm 0.0cm 0.0cm, clip]{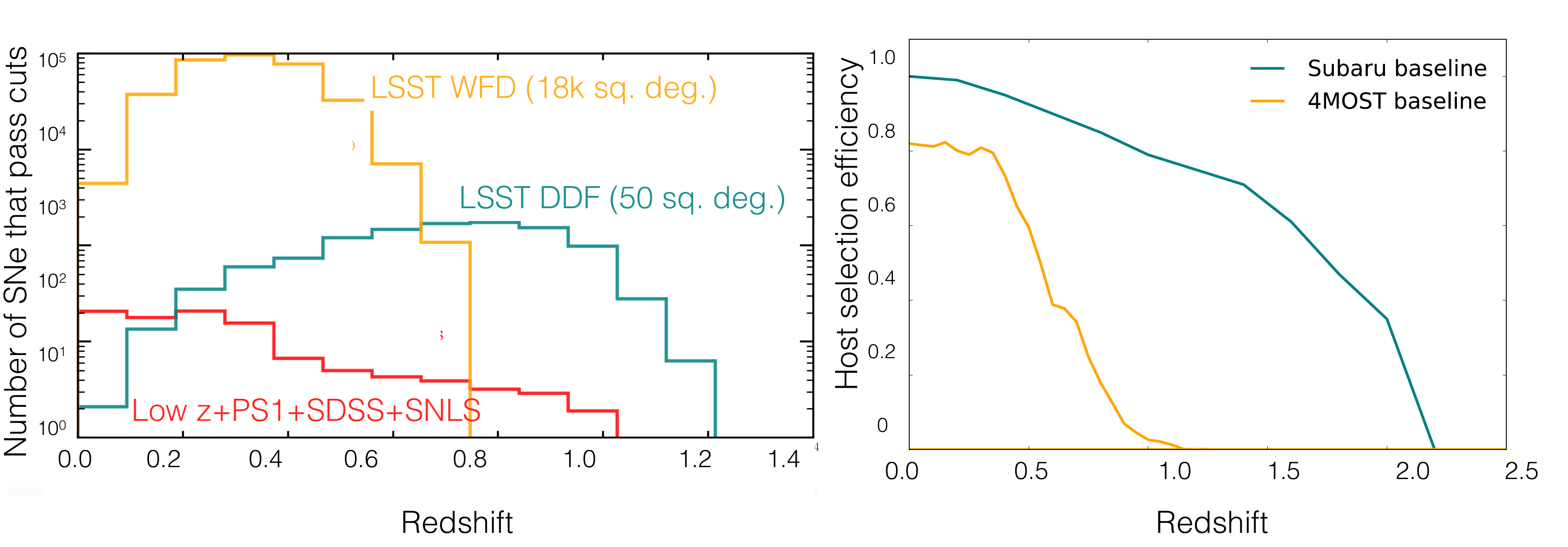}
\caption{\footnotesize \textit{Left panel:} The number of high-quality supernovae that should be observed in the wide (WFD) and deep (DDF) regions of the LSST survey. LSST will vastly increase SN samples compared to current surveys; spec-$z$'s will greatly increase the value of the SNe for cosmological analyses. \textit{Right panel:} The fraction of SNe Ia expected to have secure redshift measurements from 4MOST/TiDES observations of $i < 22$ SN hosts in the WFD region (orange curve), as well as expectations for an $i < 22.9$ Subaru/PFS sample in the DDFs (blue curve); the existence of such data is assumed in Ref.~\cite{descsrd}, and will be vital for SN Ia cosmology with LSST.  \vspace{-0.1in}}
 \label{fig:sn_specsize}
 \end{center}
 \end{figure}

While the smaller area of the deep fields ensures that host redshifts can be obtained for a higher fraction of all supernova hosts than in the WFD, the wide field will yield a much larger number of supernovae in total. 
This sample of hundreds of thousands of SNe will revolutionize cosmological analyses and enable extensive studies of systematics. The improvement is illustrated in the left panel of Figure~\ref{fig:sn_specsize}, which shows redshift histograms for the best-characterized Type Ia SNe in the DDF and WFD, as compared to the current sample of known SNe. The LSST WFD survey will increase the number of supernovae dramatically in the intermediate redshift range, complementing the deep sample from the DDF.  The 4MOST/TIDES program will target supernovae and their hosts for spectroscopy across the WFD footprint, but will only run for the first half of the LSST survey \cite{tides}.


The distribution of supernova host redshift measurements will depend on the allocation of spectroscopic resources from ground-based telescopes.  We have tested the impact on the Dark Energy Task Force Figure of Merit (FoM) from altering the fiducial redshift efficiencies from \cite{descsrd}, which are shown in Figure~\ref{fig:sn_specsize},
either by changing the total number of hours and simply rescaling the amplitude of the selection function, or by modifying the distribution to disfavor fainter hosts at higher $z$. 
From these tests we find that the DETF FoM is most sensitive to  the number of supernovae at high redshift, and hence reductions in the time allocated to the DDFs have the greatest effect.
However, high redshift completeness can be achieved across a broad range of galaxy populations for low-$z$ SNe in WFD; that will likely be impossible for the higher-$Z$ DDF sample.  This will make wide-area spectroscopy vital for investigating systematic uncertainties related to host galaxy properties, which may be a limiting factor in LSST SN cosmology.
%

\section{Recommendations}


LSST cosmological constraints can be significantly improved by additional wide-field spectroscopy of several types:
\begin{itemize}
  \item Increasing the area in the southern hemisphere with DESI-like target selection of galaxies and quasars would improve cross-correlation calibration of photo-$z$'s and other cross-correlation science. This could be achieved by performing additional surveys with the 4MOST instrument covering thousands of square degrees.
  
\item Denser and/or higher-redshift sampling of galaxies over thousands of square degrees (e.g., in the Northern part of the LSST survey footprint) could significantly improve constraints on theories of modified gravity and models of intrinsic alignments for weak lensing studies. Such a project could be pursued efficiently with the DESI spectrograph, PFS at Subaru, or the Maunakea Spectroscopic Explorer (MSE, \cite{mse}). 
  
\item Spectroscopic follow-up for strong lens systems and SNe/hosts will still be needed after 4MOST/TiDES observations are finished in 2027 \cite{tides}.  This could be pursued via an extension of the 4MOST survey or by using PFS (for deep drilling field surveys), DESI, or MSE, though the latter facilities  cannot reach the southern end of the LSST footprint.\footnote{Total observing times needed to complete the DDF supernova host survey described in \cite{kavli} on various instruments are provided at \url{http://d-scholarship.pitt.edu/id/eprint/36041}.}  Due to the rarity of these objects, this work is best pursued in combination with other projects that will use the majority of available fibers.

\item We have focused above on existing and proposed projects. However, the optimal solution for wide-field spectroscopy spanning the full LSST footprint would be a highly-multiplexed, wide field-of-view instrument in the South, ideally on an aperture at least as large as LSST's; unfortunately no such capability exists in the US OIR system. Such a facility could improve LSST cosmology while pursuing other projects with different fibers (e.g., wide-field surveys of stars in the Milky Way halo or intensive surveys of variable and transient sources in the LSST DDFs, as described in \cite{kavli}).  Such a facility could lead to advances across a broad range of astrophysics, expanding the legacy of LSST.
\end{itemize}


\newpage
\begin{center}
{\large\bf Acknowledgements}    
    
\end{center}

The LSST Dark Energy Science Collaboration acknowledges ongoing support from the Institut National de Physique Nucl\'eaire et de Physique des Particules in France; the Science \& Technology Facilities Council in the United Kingdom; and the Department of Energy, the National Science Foundation, and the LSST Corporation in the United States.  DESC uses resources of the IN2P3 Computing Center (CC-IN2P3--Lyon/Villeurbanne - France) funded by the Centre National de la Recherche Scientifique; the National Energy Research Scientific Computing Center, a DOE Office of Science User Facility supported by the Office of Science of the U.S.\ Department of Energy under Contract No.\ DE-AC02-05CH11231; STFC DiRAC HPC Facilities, funded by UK BIS National E-infrastructure capital grants; and the UK particle physics grid, supported by the GridPP Collaboration.  This work was performed in part under DOE Contract DE-AC02-76SF00515.

\newpage
\bibliographystyle{unsrt_truncate}
\bibliography{main}




\end{document}